\documentclass[showpacs,twocolumn,aps]{revtex4}
\usepackage{amssymb}
\usepackage{amsmath}
\usepackage{graphicx}
\usepackage{lscape}
\begin{document}
\title{Centrality dependence of forward-backward multiplicity correlation \\
in Au+Au collisions at $\sqrt{s_{NN}}$=200 GeV}
\author{Yu-Liang Yan$^{1}$, Dai-Mei Zhou$^{2}$, Bao-Guo Dong$^{1,3}$, Xiao-Mei
Li$^{1}$, Hai-Liang Ma$^{1}$, Ben-Hao Sa$^{1,2,4}$\footnote{Corresponding
author: sabh@ciae.ac.cn}}
\address{1 China Institute of Atomic Energy, P.O. Box 275(18), Beijing 102413,
China \\
2 Institute of Particle Physics, Huazhong Normal University, Wuhan 430079,
China\\
3 Center of Theoretical Nuclear Physics, National Laboratory of Heavy Ion
Accelerator of Lanzhou, Lanzhou 730000,China\\
4 China Center of Advanced Science and Technology, World Laboratory,
P. O. Box 8730 Beijing 100080, China}

\begin{abstract}
We have studied the centrality dependence of charged particle
forward-backward multiplicity correlation strength in Au+Au
collisions at $\sqrt{s_{NN}}$=200 GeV with a parton and hadron
cascade model, PACIAE, based on PYTHIA. The calculated results are
compared with the STAR data. The experimentally observed correlation
strength characters: (1) the approximately flat pseudorapidity
dependence in central collisions and (2) the monotonous decrease
with decreasing centrality are well reproduced. However the
theoretical results are larger than the STAR data for the peripheral
collisions. A discussion is given for the comparison among the
different models and STAR data. A prediction for the forward-backward
multiplicity correlation in Pb+Pb collisions at $\sqrt{s_{NN}}$=5500
GeV is also given.
\end{abstract}

\maketitle

\section{INTRODUCTION}
The study of fluctuations and correlations has been suggested as a
useful means for revealing the mechanism of particle production and
Quark-Gluon-Plasma (QGP) formation in relativistic heavy ion
collisions \cite{hwa2,naya}. Correlations and fluctuations of the
thermodynamic quantities and/or the produced particle distributions
may be significantly altered when the system undergoes phase
transition from hadronic matter to quark-gluon matter because of the
very different degrees of freedom between two matters.

The experimental study of fluctuations and correlations becomes a
hot topic in relativistic heavy ion collisions with the availability
of high multiplicity event-by-event measurements at the CERN-SPS and
BNL-RHIC. An abundant experimental data have been reported
\cite{star2,phen2,phob,star} where a lot of new physics arise
and are urgent to be studied. A lot of theoretical investigations
have been reported as well \cite{paja,hwa1,brog,yan,fu,bzda,konc,yan2}.

Recently STAR collaboration have measured the charged particle
forward-backward (FB) multiplicity correlation strength $b$ in a
given centrality bin size at different centralities from the most
central to peripheral Au+Au collisions at $\sqrt{s_{NN}}$=200 GeV
\cite{star}. The outstanding features of STAR data are:
\begin{enumerate}
\item In most central collisions, the correlation strength is
approximately independent of the distance between the centers of
forward and backward pseudorapidity bins, $\Delta \eta$. \item The
correlation strength monotonously decreases with decreasing
centrality. \item In the peripheral collisions, the correlation
strength approaches to an exponential function of $\Delta \eta$.
\end{enumerate}
A lot of theoretical interests \cite{fu,bzda,konc,yan2} has been stimulated.

The wounded nucleon model was used in \cite{bzda} to study the
correlation strength, the first two characters of STAR data were
reproduced, but the third one was not. In Ref. \cite{konc}, the
Glauber Monte Carlo code (GMC) with a ``toy" wounded-nucleon model
and the Hadron-String Dynamics (HSD) transport approach have been
used to analyze the STAR data. They used three different
centrality determinations: the impact parameter $b_i$, the number
of participant (wounded) nucleons $N_{part}$, and the charged
particle multiplicity $N_{ch}$ in midrapidity $\vert\eta \vert<1$.
The first two characters of STAR data can be reproduced by the
$N_{part}$ and $N_{ch}$ centrality determinations, while the third
character can not.

We have used a parton and hadron cascade model, PACIAE, to
investigate the centrality bin size dependence of charged particle
multiplicity correlation in most central (5, 0-5, and 0-10\%) Au+Au
collisions at $\sqrt{s_{ \rm{NN} }}$=200 GeV \cite{yan2}. It
turned out that the correlation strength increases with increasing
bin size. In \cite{yan2} we follow \cite{cape} defining the
charged particle FB multiplicity correlation strength $b$ as
\begin{equation}
\ b =\frac{\langle n_fn_b\rangle - \langle n_f\rangle \langle
n_b\rangle}{\langle n_f^2\rangle - \langle n_f\rangle^2} =
\frac{{\rm {cov}}(n_f,n_b)}{{\rm{var}}(n_f)},
\end{equation}
where $n_f$ and $n_b$ are, respectively, the number of charged
particles in forward and backward pseudorapidity bins defined
relatively and symmetrically to a given pseudorapidity $\eta$.
$\langle n_f\rangle$ refers to the mean value of $n_f$ for
instance. cov($n_f$,$n_b$) and var($n_f$) are the FB
multiplicity covariance and forward multiplicity variance,
respectively.

In this paper we use the PACIAE model to study the charged particle
FB multiplicity correlation strength $b$ in a given centrality bin
size at different centralities from the most central to peripheral
Au+Au collisions at $\sqrt{s_{NN}}$=200 GeV. A discussion is given
for the comparison among the models and STAR data as well as the
STAR's convention of different centrality determination for different
measured $\Delta \eta$ points. A prediction for the forward-backward
multiplicity correlation in Pb+Pb collisions at $\sqrt{s_{NN}}$=5500
GeV is also given.

\section {THE PACIAE MODEL}
The parton and hadron cascade model, PACIAE \cite{sa}, is based on
PYTHIA \cite{soj2} which is a model for hadron-hadron ($hh$)
collisions. The PACIAE model is composed of four stages:
the parton initialization, parton evolution (rescattering),
hadronization, and hadron evolution (rescattering).

1. PARTON INITIALIZATION: In the PACIAE model, a nucleus-nucleus
collision is decomposed into the nucleon-nucleon collisions based on
the collision geometry. A nucleon-nucleon ($NN$) collision is
described with the PYTHIA model, where a $NN$ (hadron-hadron, $hh$)
collision is decomposed into the parton-parton collisions. The hard
parton-parton collision is described by the lowest-leading-order
(LO) pQCD parton-parton cross section \cite{comb} with modification
of parton distribution function in the nucleon. And the soft
parton-parton interaction is considered empirically. The semihard,
between hard and soft, QCD $2\rightarrow 2$ processes are also
involved in the PYTHIA (PACIAE) model. Because of the initial- and
final-state QCD radiation added to the parton-parton collision
processes, the PYTHIA (PACIAE) model generates a partonic multijet
event for a $NN$ ($hh$) collision. That is followed, in the PYTHIA
model, by the string-based fragmentation scheme (Lund string model
and/or Independent Fragmentation model), thus a hadronic final state
is reached for a $NN$ ($hh$) collision. However, in the PACIAE model
the above fragmentation is switched off temporarily, so the result
is a partonic multijet event (composed of quark pairs, diquark pairs
and gluons) instead of a hadronic final state. If the diquarks
(anti-diquarks) are split forcibly into quarks (anti-quarks)
randomly, the consequence of a $NN$ ($hh$) collision is its initial
partonic state composed of quarks, anti-quarks, and gluons.

2. PARTON EVOLUTION: The next stage in the PACIAE model is the
parton evolution (parton rescattering). Here the $2\rightarrow 2$
LO-pQCD differential cross sections \cite{comb} are employed. The
differential cross section of a subprocess $ij\rightarrow kl$ reads
\begin{equation}
\frac{d\sigma_{ij\rightarrow
kl}}{d\hat{t}}=K\frac{\pi\alpha_s^2}{\hat{s}}\sum_{ij\rightarrow
kl},
\end{equation}
where the factor $K$ is introduced considering the higher order pQCD
and non-perturbative QCD corrections as usual, $\alpha_s$ stands for
the strong (running) coupling constant. Taking the process $q_1q_2
\rightarrow q_1q_2$ as an example one has
\begin{equation}
\sum_{q_1q_2\rightarrow
q_1q_2}=\frac{4}{9}\frac{\hat{s}^2+\hat{u}^2}{\hat{t}^2},
\label{eq3}
\end{equation}
where $\hat{s}$, $\hat{t}$, and $\hat{u}$ are the Mandelstam
variables. Since it diverges at $\hat{t}$=0 one has to regularize it
with the parton color screen mass $\mu$
\begin{equation}
\sum_{q_1q_2\rightarrow
q_1q_2}=\frac{4}{9}\frac{\hat{s}^2+\hat{u}^2}{(\hat{t}-\mu^2)^2}.
\end{equation}

The total cross section of the parton collision, $i+j$, then reads
\begin{equation}
\sigma_{ij}(\hat{s})=\sum_{k,l}\int_{-\hat{s}}^{0}d\hat{t}
\frac{d\sigma_{ij\to kl}}{d\hat{t}}.
\end{equation}
With the total and differential cross sections above the parton
evolution (parton rescattering) can be simulated by the Monte Carlo
method.

3. HADRONIZATION: The parton evolution stage is followed by the
hadronization at the moment of partonic freeze-out (no any more
parton collision). In the PACIAE model, the Lund string
fragmentation model and phenomenological coalescence model are
supplied for the hadronization of partons after rescattering. The
Lund string fragmentation model is adopted in this paper. We refer
to \cite{sa} for the details of the hadronization stage.

4. HADRON EVOLUTION: After hadronization the rescattering among
produced hadrons is dealt with the usual two-body collision model.
Only the rescattering among $\pi, k, p, n, \rho (\omega), \Delta,
\Lambda, \Sigma, \Xi, \Omega, J/\Psi$ and their antiparticles are
considered in the calculations. An isospin averaged parametrization
formula is used for the $hh$ cross section \cite{koch,bald}. We also
provide an option for the constant total, elastic, and inelastic
cross sections ($\sigma_{\rm{tot}}^{NN}=40$~mb, $\sigma_{\rm{tot}
}^{\pi N}=25$~mb, $\sigma_{\rm{tot}}^{kN}=35$~mb,
$\sigma_{\rm{tot}}^{\pi \pi}=10$~mb) and the assumed ratio of
inelastic to total cross section of 0.85. More details of hadronic
rescattering can see \cite{sa1}.

\section {CALCULATIONS AND RESULTS}
In our calculations the default values given in the PYTHIA model
are adopted for all model parameters except the parameters $K$ and
$b_s$ (in the Lund string fragmentation function). $K$=3 is
assumed and $b_s$=6 is tuned to the PHOBOS data of charged
particle multiplicity in 0-6\% most central Au+Au collisions at
$\sqrt{s_{NN}}$=200 GeV \cite{phob2}, as shown in Tab.~\ref{mul}.

\begin{table}[htbp]
\caption{Total charged particle multiplicity in three $\eta$
fiducial ranges in 0-6\%
     most central Au+Au collisions at $\sqrt{s_{\rm{NN}}}$=200 GeV.}
\begin{tabular}{cccc}
\hline\hline
         & $N_{\rm{ch}}(|\eta|< 4.7)$  & $N_{\rm{ch}}(|\eta|< 5.4) $ &
$N_{\rm{ch}}$(total) \\
\hline
  PHOBOS$^a$ & 4810 $\pm$ 240  & 4960 $\pm$ 250 &  5060 $\pm$ 250                 \\
  PACIAE     & 4819  & 4983    &  5100    \\

\hline\hline
\multicolumn{4}{l}{$^a$ The experimental data are taken from \cite{phob2}.}\\
\end{tabular}
\label{mul}
\end{table}

In the theoretical calculation it is convenient to define the
centrality by impact parameter $b_i$. The mapping relation between
centrality definition in theory and experiment
\begin{equation}
b_i=\sqrt{g}b_i^{\rm {max}},\qquad b_i^{\rm {max}}=R_A + R_B,
\label{imp}
\end{equation}
is introduced \cite{sa2}. In the above equation, $g$ stands for
the geometrical (total) cross section percentage (or charged multiplicity
percentage) used in the experimental determination of centrality.
$R_A=1.12A^{1/3}+0.45$ fm is the radius of nucleus $A$. We have also used
the centrality determination based on the charged multiplicity $N_{ch}$ in
the central pseudorapidity window $\vert\eta \vert<1$ to study the
FB multiplicity correlation strength as a function of $\Delta \eta$.

\begin{figure}[htbp]
\includegraphics[width=8cm,angle=0]{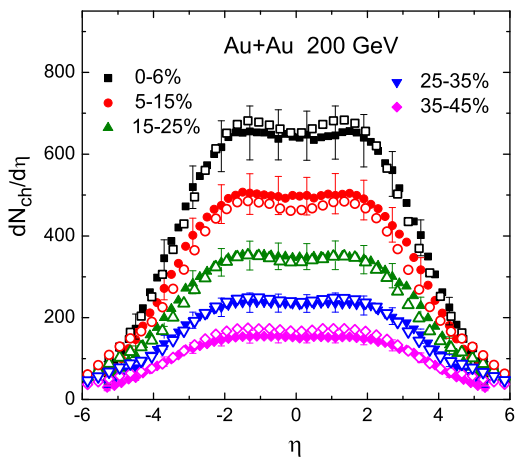}
\caption{(Color online) The charged particle pseudorapidity
distribution for the specified centrality bin in Au+Au collisions at
$\sqrt{s_{NN}}$=200 GeV. The solid and open symbols are the PHOBOS
data \cite{phob2} and PACIAE results, respectively.} \label{bau2_1}
\end{figure}

We compare the calculated charged particle pseudorapidity
distribution in specified centrality bin in Au+Au collisions at
$\sqrt{s_{NN}}$=200 GeV with the corresponding PHOBOS data
\cite{phob2} in Fig.~1. Here the theoretical centrality are
defined by impact parameter $b_i$ via Eq.(\ref{imp}). One sees
that the PHOBOS data are well reproduced.

\begin{figure}[htbp]
\includegraphics[width=8cm,angle=0]{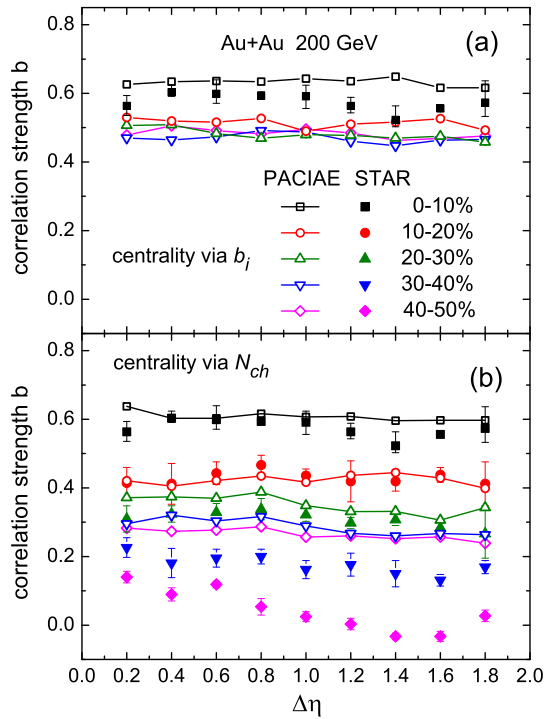}
\caption{(Color online) The FB multiplicity correlation strength
$b$ in Au+Au collisions at $\sqrt{s_{NN}}$=200 GeV for centrality
determination of (a) impact parameter $b_i$ and (b) charged particle
multiplicity $N_{ch}$ ($\vert\eta \vert<1$). The open symbols with
line and solid symbols are the PACIAE results and STAR data
\cite{star}, respectively.} \label{bau2_2}
\end{figure}

\begin{figure}[htbp]
\includegraphics[width=8cm,angle=0]{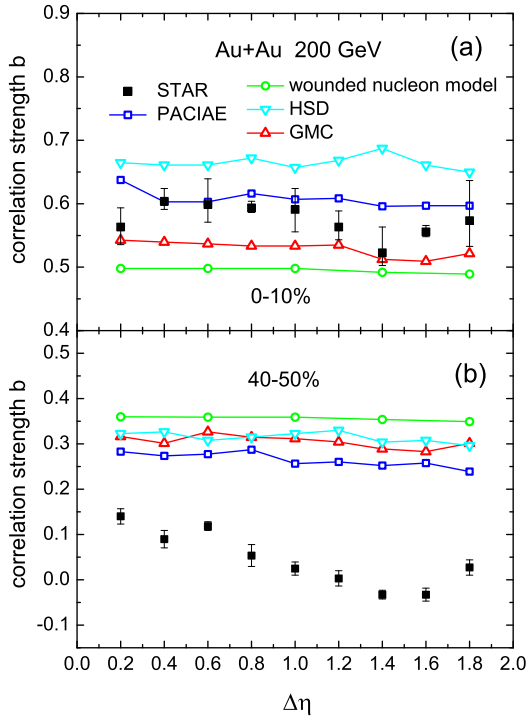}
\caption{(Color online) The FB multiplicity correlation strength
$b$ calculated by different models are compared with the STAR
data\cite{star} for (a) 0-10\% and (b) 40-50\% central Au+Au
collisions at $\sqrt{s_{NN}}$=200 GeV.} \label{bau2_3}
\end{figure}

The PACIAE model results of FB multiplicity correlation strength
are shown in Fig.~2 (a) and (b) for the centrality determinations
of the impact parameter $b_i$ and charged multiplicity $N_{ch}$
($\vert\eta \vert<1$), respectively. One sees in Fig.~2 (a) that
the theoretical results can reproduce the STAR data for 0-10\%
most central collisions. However the theoretical results are all
higher than the corresponding STAR data for the 10-20\%, 20-30\%,
30-40\%, and 40-50\% most central collisions and those theoretical
results are closed to each other. That is because the FB
multiplicity correlation in Au+Au collisions is mainly the
statistical correlation steaming from the multiplicity fluctuation
\cite{yan2}. And the multiplicity fluctuation in those
centralities defined by impact parameter are similar to each
other.

Figure~2 (b) gives the FB correlation strength calculated with the
charged particle multiplicity ($\vert\eta \vert<1$) centrality
determination together with the STAR data. We see in this panel
that the STAR data are well reproduced by the PACIAE model for
the 0-10\%, 10-20\%, and 20-30\% central collisions. However, the
PACIAE results are higher than the corresponding STAR data for
30-40\%, and 40-50\% central collisions. The correlation strength
approaches an exponential function of $\Delta \eta$ observed by
STAR in 40-50\% central collisions can not be reproduced
especially.

In Fig.~3 we compare the FB multiplicity correlation strength as a
function of $\Delta \eta$ calculated by the wounded nucleon model
\cite{bzda}, GMC code with a ``toy" wounded-nucleon model, HSD
transport approach \cite{konc}, and PACIAE model with the STAR data
for the 0-10\% (panel(a)) and 40-50\% central collisions (panel (b)).
We see in Fig.~3 (a) that all of the four models can nearly reproduce
the STAR data for 0-10\% central collisions and the PACIAE model is
somewhat better than others. However all of the model results are higher
than the STAR data for the 40-50\% central collisions, therefore can not
reproduce $b$ as an exponential function of $\Delta \eta$ observed by
STAR as shown in Fig.~3 (b).

\begin{figure}[htbp]
\includegraphics[width=8cm,angle=0]{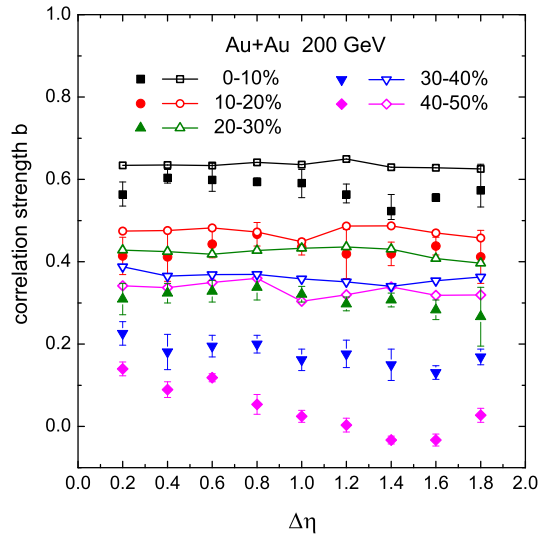}
\caption{(Color online) The same as Fig.~2 (b) but the PACIAE
results are calculated with STAR's convention of different centrality
determination for different measured $\Delta \eta$ points (see text for
the details).} \label{bau2_4}
\end{figure}

We have noted the mentions in \cite{star} that ``To avoid a bias in
the FB correlation measurements, care was taken to use different
pseudorapidity selection for the centrality determination which is
also based on multiplicity. Therefore, the centrality determination
for the FB correlation strength for $\Delta \eta$=0.2, 0.4, and 0.6
is based on the multiplicity in $0.5<\vert\eta \vert<1.0$, while for
$\Delta \eta$=1.2, 1.4, 1.6, and 1.8, the centrality is obtained
from $\vert\eta\vert<0.5$. For $\Delta \eta$=0.8 and 1.0, the sum of
multiplicities from $\vert\eta \vert<0.3$ and $0.8<\vert\eta
\vert<1.0$ is used for the centrality determination." So we follow
STAR's centrality determination convention to repeat all of the PACIAE
calculations and draw Fig.~4 with these results. Comparing Fig.~4
with Fig.~2 (b) we see this complicated centrality determination
does not improve but even worsens the agreement between experiment
and theory. That means STAR's centrality determination convention may
be needed for the experimental measurement of FB correlation strength
but not for the theoretical calculation. In this kind of theoretical
calculations, each definite centrality curve in Fig.~4 is composed
of $b$ calculated at different $\Delta \eta$ with different centrality
determination, such kind of curve is not reasonable in theoretical
physics.

\begin{figure}[htbp]
\includegraphics[width=8cm,angle=0]{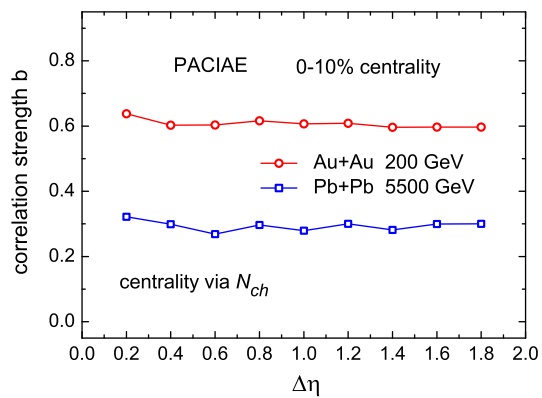}
\caption{(Color online) The PACIAE calculated FB multiplicity
correlation strength $b$ in Pb+Pb collisions at $\sqrt{s_{NN}}$=5500
GeV with centrality determination of charged particle multiplicity
$N_{ch}$ ($\vert\eta \vert<1$).} \label{bau2_5}
\end{figure}

Figure~5 gives the PACIAE model prediction for the charged particle
FB multiplicity correlation strength in Pb+Pb collisions at $\sqrt
{s_{NN}}$=5500 GeV. We can see that the FB multiplicity correlation
strength in Pb+Pb collisions is smaller than in Au+Au collisions. One
reason may be that the multiplicity in Pb+Pb collisions at LHC energy
is much larger than in Au+Au collisions at RHIC energy (the $N_{ch}/
d\eta$ at mid-rapidity, $|\eta|<$0.5, is about 600 in 0-10\% most central
Au+Au collisions at RHIC energy but it is around 1200 in 0-10\% most
central Pb+Pb collisions at LHC energy \cite{sa3}). This observation is
similar to the report that the elliptic flow parameter $v_2$ in Pb+Pb
collisions at LHC energy is significantly smaller than in Au+Au collisions
at RHIC energy \cite{eyyu}. That is attributed to the fact that the hard
process is more influential at the LHC energy than RHIC energy in \cite{eyyu}.
Whether the competition between the hard and soft processes is also the
reason of the FB multiplicity correlation decreasing from RHIC energy to
LHC energy is beyond this paper scope, and it would be studied later.

\section {CONCLUSION}
We have used a parton and hadron cascade model, PACIAE, to study the
centrality dependence of charged particle FB multiplicity
correlation strength in 0-10\%, 10-20\%, 20-30\%, 30-40\%, and
40-50\% central Au+Au collisions at $\sqrt{s_{NN}}$=200 GeV. For the
0-10\%, 10-20\%, and 20-30\% central collisions, the STAR data are well
reproduced. The STAR observed characters of (1) $b$ as a function of
$\Delta \eta$ is approximately flat for central collisions and (2) $b$
decreases with decreasing centrality are reproduced as well. However the
PACIAE results are higher than the STAR data for the 30-40\% and 40-50\%
central collisions and can not obtain $b$ as an exponential function of
$\Delta \eta$ for the 40-50\% central collisions, especially.

It turned out that the PACIAE model is somewhat better than the wounded
nucleon model, the GMC code with a ``toy" wounded-nucleon model, or the
HSD transport model in comparing with the STAR correlation data. However
all the models can not reproduce $b$ as an exponential function of
$\Delta \eta$ in the 40-50\% central collisions observed by STAR. That
should be studied further.

The PACIAE calculations are repeated using the STAR's centrality
determination convention mentioned above. The results not improve but
even worsens the agreement between theory and experiment. This means
STAR's centrality determination convention may be needed for the
experimental measurement but not for the theoretical calculations.
Because in such theoretical calculations, each definite centrality curve
is composed of $b$ calculated at different $\Delta \eta$ with different
centrality determination, this kind of curve is not reasonable in
theoretical physics.

A prediction for the charged particle FB multiplicity correlation strength
in 0-10\% Pb+Pb collisions at $\sqrt{s_ {NN}}$=5500 GeV is also given. The
charged particle FB multiplicity correlation strength in Pb+Pb collisions
at LHC energy is much smaller than in Au+Au collisions at RHIC energy.
The further study is out of present paper scope and has to be investigated
in another paper.

\begin{center} {ACKNOWLEDGMENT} \end{center}
Finally, the financial support from NSFC (10635020, 10605040,
10705012, 10475032, 10975062, and 10875174) in China is
acknowledged.

\end{document}